\documentclass[10pt,a4paper,twoside]{article}

\usepackage{a4}
\usepackage{amsfonts}
\usepackage{amssymb}
\usepackage{epsfig}
\usepackage{psfig}

\newcommand{\beqn}{\begin{eqnarray}}
\newcommand{\eeqn}{\end{eqnarray}}
\newcommand{\be}{\begin{equation}}
\newcommand{\ee}{\end{equation}}
\newcommand{\non}{\nonumber \\}
 
\newcommand{\Tr}{\mbox{Tr}}

\newcommand{\sz}{\scriptsize}
\newcommand{\ns}{{\mbox{\sz NSNS}}}
\newcommand{\rr}{{\mbox{\sz RR}}}
\newcommand{\utw}{{\mbox{\sz U}}}
\newcommand{\tw}{{\mbox{\sz T}}}
     
\begin{document} 

\title{
\begin{flushright} {\small HUB-EP-99/30} \end{flushright}
\vspace{1cm}
{\bf D-brane Spectra of Nonsupersymmetric, Asymmetric 
Orbifolds and Nonperturbative Contributions to the 
Cosmological Constant}}
\author{} 
\date{}

\maketitle

\begin{center}
{\bf Boris K\"ors}\footnote{Email: Koers@Physik.HU-Berlin.De}
\hspace{-0.11cm}\\
\vspace{0.5cm}
Humboldt Universit\"at zu Berlin \\
{\small{Institut f\"ur Physik, Invalidenstr. 110, 10115 Berlin, Germany}} 
\vspace{2cm}
\end{center}

\begin{abstract}
We study nonperturbative aspects of asymmetric orbifolds of 
type IIA, focussing on models that allow a dual perturbative 
heterotic description. In particular we derive the boundary 
states that describe the nonsupersymmetric D-branes of the untwisted sector 
and their zero mode spectra. These we use to demonstrate, how some 
special non BPS multiplets are identified under the duality map, 
and give some indications, 
how the mismatch of bosons and fermions in the perturbative heterotic 
spectrum is to be interpreted in terms of the nonperturbative degrees 
of freedom on the type IIA side.

\end{abstract}

\thispagestyle{empty}
\clearpage

\section{Introduction}

Quite recently certain properties 
of asymmetric orbifolds of type II and also type I string  
theories have been discussed, and they were found to 
display several phenomenologically attractive features, among them the 
opportunity to have a vanishing contribution to the 
cosmological constant $\Lambda_{\mbox{\sz c}}$, at least to 
leading order in perturbation theory, without supersymmetry but with a 
degeneracy of bosons and fermions at any mass level. 
Such models of type II on a torus $T^6$ divided out by an asymmetrically and freely 
acting orbifold group were constructed by Kachru, 
Kumar and Silverstein (KKS) in \cite{hepth9807076,hepth9810129}. 
They argued, that the contribution of the 
vacuum fluctuations to $\Lambda_{\mbox{\sz c}}$ 
might in fact be vanishing at all orders in perturbation theory. 
Furthermore by a self-duality argument even nonperturbative contributions 
are believed to be absent \cite{hepth9808056}. The applicability of  
these dualities in nonsupersymmetric but freely acting orbifolds derives from  
the well known but still heuristic adiabatic argument 
\cite{hepth9507050}, which appears to be less reliable than in any supersymmetric 
case. \\

In a slight variation of the original KKS model, 
two shifts exchanged and one circle decompactified but 
with still keeping the 
above mentioned perturbative features, Harvey was able to explicitly 
compute the nonperturbative contributions to $\Lambda_{\mbox{\sz c}}$ in a dual 
perturbative heterotic model \cite{hepth9807213}. 
By splitting the orbifold group in a suitable manner, one can obtain a 
symmetric spatial reflection in an intermediate step and employ the IIA on 
K3$\times S^1$ to heterotic on $T^5$ duality. Such one obtains 
a dual model of the asymmetric IIA orbifold. Its partition function 
displayes a nonvanishing but exponentially 
suppressed contribution. This raises the question, if there 
is a way to identify the nonperturbative degrees of freedom of the 
IIA theory, that lead to the mismatch of bosons and fermions on the 
heterotic side. They should be describable as excitations of a IIA soliton 
compactified on K3$\times S^1$ and further divided out by the asymmetric orbifold 
group, or of some wrapped D-brane. As we are dealing 
with nonsupersymmetric orbifolds, the cycles will not be holomorphic 
and the corresponding states cannot be 
deduced from any BPS argument. We shall employ the boundary state formalism 
to identify the non BPS D-brane spectra. \\

Of course, one has to 
worry about the question, in which situations there is any hope to extrapolate 
the perturbative open string analysis of strings ending on the respective 
brane to strong coupling, which is necessary to understand the dual spectra 
of the heterotic model, while the usual BPS argument is absent. A possible 
answer is to focus on states which are perturbatively the lightest to carry 
some particular conserved charge or other quantum number. These states will 
dominate the dynamics, if uncharged states are prohibited by the conservation law, 
and therefore should be present on both sides of the duality, if gauge 
invariance is respected. Such cases 
have been studied recently in a number of different dualities, including in 
particular dual 
orbifolds and orientifolds of IIB \cite{hepth9803194,hepth9805019,
hepth9806155}, the type I to $SO(32)$ heterotic \cite{hepth9808141} as 
well as the IIA on K3 to heterotic on $T^4$ dualities \cite{hepth9812031,hepth9901014}. 
For instance, an interpretation of a D-brane wrapping 
a nonsupersymmetric cycle in terms of a superposition 
of a brane and an anti-brane was given, that involved the appearance of tachyons in 
the spectra of open strings, more precisely in the Chan-Paton 
sector of those strings with one end on the brane and one on the 
anti-brane. By the application of the formerly invoked tachyon condensation 
\cite{hepth9808141,hepth9805170} these potentially instable states can, 
at least for appropriate values of the radii of the compact dimensions, 
have a condensate of the tachyon field on their world volume, that neutralizes 
their energy density everywhere except in 
some lower dimensional region. Such the terminology, that some tachyonic 
pairs of BPS branes can condensate into a single non BPS brane, defined 
by the vortex of the tachyon field, was established. This mechanism was 
justified by a strict CFT treatment for the case of a D1-brane and 
-anti-brane pair in type I on $\mathbb{R}^{1,8} \times S^1$ 
\cite{hepth9808141,hepth9903123}, analyzing the 
marginal deformation that comes from the massless tachyon at a special, 
critical radius. It corresponds to deforming the 1+1 dimensional 
CFT on the brane pair from a theory 
without to one with a tachyon condensate. We shall take the attitude, that this 
analysis can be carried over to the cases we are discussing presently 
without any qualitative modification. \\

In the following we analyze the asymmetric orbifolds and compare some of 
their charged states, given by perturbative excitations of 
the heterotic string and of D-branes of type IIA, both subject to the orbifold 
projection. This allows to reveal some indication for the origin of the 
nonperturbative Bose Fermi mismatch of Harvey's example. The key 
to do this is a study of lightest charged states that can be reliably 
identified on both sides of the duality and also violate the degeneracy of 
bosons and fermions when subject to the orbifold projection. 
The further proceeding will be as follows: In section \ref{ch2} we shall 
use the boundary state formalism to derive the possible BPS and non BPS Branes 
of IIA theory on the $\mathbb{R}^{1,9}/\mathbb{Z}_2 ={\mbox{K3}} \times \mathbb{R}^{1,5}$ 
orbifold and their zero mode spectra. In section \ref{ch3} we 
review some aspects of the duality IIA on K3$\times S^1$ to the heterotic 
string on $T^5$, in particular collect all BPS and some non BPS charges  
and recall, how the asymmetric elements of the orbifold group 
act on both sides. We shall find, that the heterotic Bose Fermi mismatch is indeed 
present in certain multiplets which are the lightest states to carry some charge under 
the gauge group. Finally in section \ref{ch4} we put both together and 
show, how this can be recovered on a heuristic level in terms of 
the zero modes of D-branes on K3$\times S^1$. We close with a summary.

\section{IIA on a $\mathbb{Z}_2$ orbifold}
\label{ch2}

We are going to construct all the possible D-brane states in the 
$\mathbb{Z}_2$ orbifold of IIA on flat 
space-time, given by dividing out by a reflection $I_4$ in four 
directions, say $x_1,...,x_4$. 
These branes carry over to the compactified theory on K3 at its 
orbifold limit $T^4/{I_4}$, only modified by zero mode terms. 
The lightest excitations of orbifold invariant superpositions of such branes 
will be compared to perturbative states of the heterotic asymmetric orbifold later on. 

\subsection{D-brane spectra}

Examples of such states have been discussed very many times 
by different authors, and we only like to collect the facts we shall 
need, putting them into a suitably systematic order. The states will be 
identified using the boundary state formalism, which, to our 
knowledge, has been invented in \cite{CNLY1988,PC1987}, and was used 
and reviewed by a couple of authors in the recent past. Such we take 
it as given and stick to the conventions of \cite{hepth9805019}, 
where some of the following is to be found. For the 
oscillators
\beqn \label{bsosc}
\vert p, k,\eta \rangle = \exp \left( \sum_{n=1}^\infty{ \frac{1}{n} \left( 
\alpha^i_{-n} \tilde{\alpha}_{-n}^i - \alpha^\mu_{-n} \tilde{\alpha}_{-n}^\mu 
\right)} +i\eta \sum_{r>0}^\infty{ \left( \psi_{-r}^i \tilde{\psi}_{-r}^i - 
 \psi_{-r}^\mu \tilde{\psi}_{-r}^\mu \right) } \right) \vert p, k,\eta 
\rangle^{(0)}_{\ns, \rr}
\eeqn
solves Dirichlet boundary conditions in the $7-p$\ $i$-directions transverse to 
the brane and Neumann conditions in the $p+1$\ $\mu$-directions along the 
brane, irrespective of the NSNS ($r \in \mathbb{Z}_+ +1/2$) 
or RR ($r \in \mathbb{Z_+}$) sector groundstate 
$\vert p, k,\eta  \rangle^{(0)}$. The exponential is obviously invariant under 
$I_4$, $\eta$ takes values $\pm 1$ to distinguish 
the different spin structures, and $k$ is the momentum eigenvalue. 
The use of light-cone gauge  
has the advantage of avoiding to write out ghost contributions, but of course 
it obscures covariance at several instances unavoidably. The groundstates are 
defined to satisfy the additional constraints coming from the zero modes. 
To obtain a localized state, an eigenstate of the 
position operator, one has to integrate over 
the transverse momenta $k^\perp$ in the Dirichlet directions:
\beqn \label{momint}
\vert p,\eta \rangle \equiv \delta^{(9-p)} 
\left( x- x^\perp \right) \vert p,\eta,k^\parallel=k^\perp=0 \rangle = 
\int_{-\infty}^\infty{d^{(9-p)}k^\perp\ 
\vert p,\eta,k^\parallel=0,k^\perp \rangle} .
\eeqn
One needs to introduce a relative normalization of the NSNS and 
RR components by 1 and $4i$ with respect to their ground state degeneracy and 
an over all normalization factor $N_{\mbox{\sz U}}$ for the brane volume 
Vol$_{p+1}$, which satisfies $2^5 N_{\mbox{\sz U}}^2= 
{\mbox{Vol}}_{p+1}/(2\pi )^{p+1}$. The GSO invariant untwisted 
boundary state of type IIA is  
\beqn \label{utwbs}
\vert Dp \rangle_{\mbox{\sz bulk}} & \equiv & \vert Dp \rangle^{\mbox{\sz U}} 
\equiv \vert Up \rangle_\ns \pm \vert Up \rangle_\rr , \non
\vert Up \rangle_\ns &\equiv& \vert p,+ \rangle^\utw_\ns - \vert p, 
- \rangle^\utw_\ns, 
\quad \vert Up \rangle_\rr \equiv \vert p, + \rangle^\utw_\rr + 
\vert p,- \rangle^\utw_\rr ,
\eeqn
$p$ being constrained to be even, of course. 
These are the well known BPS D-branes of IIA and all survive the orbifold 
projection. In fact only the RR component is constrained to have even 
$p$, while the NSNS might be of any dimension, which technically 
results from the fact that only the RR fields have zero modes. 
These are the bulk branes as opposed to the fractional ones 
localized at the singular fixed planes of the orbifold. 
The normalization convention is defined in a way, that the state is normalized 
to one, but carries the double amount of RR charge compared to a fractional 
boundary state which we turn to next. \\

The states of the twisted sector are necessarily localized at the fixed 
plane, i.e. their boundary states have to be integrated only over those momenta 
that are parallel to the fixed plane and transverse to the brane. The  
moding of the oscillators is changed from half-integer to integer and 
vice versa in the directions that are reflected. This leads to four (light-cone 
gauge) fermionic zero modes in each sector, the NSNS and the RR, giving 
two independent constraints on the allowed dimensions of the branes, which derive from 
\beqn
(-1)^F \vert r,\eta \rangle^\tw_\rr &=& \vert r, -\eta \rangle^\tw_\rr, \quad
(-1)^{\tilde{F}} \vert r,\eta \rangle^\tw_\rr 
= (-1)^{r+1} \vert r, -\eta \rangle^\tw_\rr, \non
(-1)^F \vert s,\eta \rangle^\tw_\ns &=& \vert s, -\eta \rangle^\tw_\ns, \quad 
(-1)^{\tilde{F}} \vert s,\eta \rangle^\tw_\ns = (-1)^{s} \vert s, -\eta 
\rangle^\tw_\ns.
\eeqn
Here $r+1$ is the total dimension of the brane in 
the orbifold fixed plane, $s$ the dimension transverse to it and $p+1=r+s+1$. 
In (\ref{momint}) one then has to replace $9-p$ by $5-r$ in the integral and 
the relative prefactors $1$ and $4i$ for the degeneracy 
of the groundstate, the dimension of the spinor, become $2$ and $2i$. Also 
the overall factor needs to be changed in accord with 
$2^3 N_{\mbox{\sz T}}^2 = 4 {\mbox{Vol}_{r+1}}/(2 \pi)^{r+1}$ 
to match the open string 
spectrum. The GSO invariant twisted sector boundary states are:
\beqn
\vert D(r,s) \rangle^{\mbox{\sz T}} 
&\equiv& \vert Ts \rangle_\ns \pm \vert Tr \rangle_\rr , \non
\vert Ts \rangle_\ns &\equiv& \vert s,+ \rangle^\tw_\ns + \vert s, - 
\rangle^\tw_\ns, 
\quad \vert Tr \rangle_\rr \equiv \vert r, + \rangle^\tw_\rr + 
\vert r,- \rangle^\tw_\rr  
\eeqn
with even $r$ and $s$ in IIA. The most general state in the D-brane 
Hilbert space has twisted and untwisted components. 
We only consider states that carry some RR charge, so there may be long
\beqn \label{long}
\vert D(p,r,s) \rangle_{\mbox{\sz long}} \equiv 
\frac{1}{2} \left( \vert Up \rangle_\ns \pm \vert Up \rangle_\rr 
\pm  \vert Ts \rangle_\ns \pm \vert Tr \rangle_\rr \right) , 
\eeqn
or short 
\beqn \label{short}
\vert D(p,r,s) \rangle_{\mbox{\sz short}} \equiv 
\frac{1}{\sqrt{2}} \left( \vert Up \rangle_\ns  \pm 
\vert Tr \rangle_\rr \right) , 
\eeqn
or entirely untwisted bulk branes as in (\ref{utwbs}). In particular, there are no states 
with odd $r$, which is of some importance for the spectrum of charges in the IIA 
orbifold on K3$\times S^1$. This analysis is nothing but a more systematic,  
slight generalization of particular examples studied previously, and all the 
boundary states of \cite{hepth9805019,hepth9806155,hepth9901014} fit into this pattern.

\subsection{Zero mode spectra and tachyonic instabilities of non BPS branes}

We also like to collect some facts about the spectra of open 
strings that end on and 
stretch between the various D-branes \cite{hepth9805019,hepth9806155}. 
Their massless excitations govern the low energy dynamics and upon quantization on the 
worldsheet of the dual string are 
to be compared to the perturbative spectrum of the heterotic string.  
Two important issues are 
the amount of supersymmetry broken by the brane, to be read off from the 
number of its fermionic zero modes, and the appearance of tachyons 
in the spectrum. This is done by computing the projector $P$ that 
has to be introduced into the open string trace to reproduce the closed 
string diagram:
\beqn
\int_0^\infty{dl\ \langle Dp \vert e^{-lH_{\mbox{\sz cl}}} \vert Dp \rangle }  
= \int_0^\infty{\frac{dt}{2t} \Tr_{\mbox{\sz NS-R}} 
\left( P\ e^{-2tH_{\mbox{\sz op}}} \right) }   .
\eeqn
Starting with the bulk brane from (\ref{utwbs}) we get the usual 
supersymmetric GSO projection, but no orbifold projection:
\beqn \label{projbulk}
P =\frac{1}{2} \left( 1+ (-1)^F \right) .
\eeqn 
This state has eight (light-cone gauge, as always) fermionic and bosonic zero 
modes, double as many fermionic 
zero modes, as one would expect for a BPS state of IIA on K3, which can be resolved by 
observing, that only half of the states are invariant under the reflection $I_4$.  
They are the Goldstones of the broken 
translation invariance and the Goldstinos of the broken 
supersymmetries which have not yet been broken by the K3. \\

Only for a boundary state $\vert Dp\rangle 
=\vert D(p,r,s) \rangle_{\mbox{\sz long}}$ one finds the orbifold and GSO projection 
expected for an invariant fractional BPS brane: 
\beqn \label{projfrac}
P= \frac{1}{4} \left( 1 + (-1)^F \right) \left( 1 + {I}_4 \right) ,
\eeqn
with all ambiguities concerning the relative signs of the components 
cancelling. The open string massless spectrum contains four bosonic and 
fermionic zero modes each, signalling a BPS 
state of the ${\cal N}=(1,1)$ supersymmetry in $d=6$. 
For a state $\vert Dp\rangle =\vert D(p,r,s) \rangle_{\mbox{\sz short}}$ 
with odd $p$ and $s$ but even $r$ the projector reads
\beqn \label{projnbps}
P = \frac{1}{2} \left( 1+ (-1)^F {I}_4 \right).
\eeqn
This spectrum has no GSO projection, four bosonic and eight fermionic   
zero modes, displaying its non 
BPS nature. For a superposition of a brane with its anti-brane, 
defined by choosing two relative signs in a superposition of two long BPS states from 
(\ref{long}) in a way to cancel the untwisted 
sector RR as well as the twisted sector NSNS charges, the projection 
\beqn \label{projpair}
P= \frac{1}{4} \left( 1 - (-1)^F \right) \left( 1 - {I}_4 \right) 
\eeqn 
in the Chan-Paton sector of strings stretching between the two 
allows for four fermionic zero modes in addition to those from the other 
Chan-Paton sector and breaks all the supersymmetry.  
The question, if any of these non BPS branes or superpositions 
will have a tachyon in the open string NS sector, depends on the orientation 
relative to the orbifold fixed plane again. 
If the non BPS brane or the brane anti-brane pair is lying entirely inside the 
orbifold fixed plane, the momenta of any open string state will be 
even under the reflection and the NS vacuum is projected out 
in (\ref{projnbps}) or (\ref{projpair}). This relative orientation 
enters into the closed string 
amplitude only through the momentum integration. If the prefactors 
$2^{(9-p)/2}$ and $2^42^{(5-r)/2}$ for the twisted and untwisted sectors 
leave a relative factor of $4$, i.e. $p-r=s=0$ and the brane 
lies entirely inside the orbifold fixed plane, there is no tachyon. 
On the other hand, if there is a tachyon, the brane system is unstable and 
the mechanism of tachyon condensation 
should take over \cite{hepth9808141,hepth9805170}. This leads to 
a presumably equivalent description in terms of another non BPS brane, 
which is assumed to be the core of the tachyon vortex. 
In the cases we consider this simply means, 
there will be a D-brane of one dimension higher or lower, 
that carries identical RR charges and is stable. Because of the naive identity
\beqn \label{gleich}
\frac{1}{2} \left( 1+(-1)^F I_4 \right) = \frac{1}{4} \left( 1+(-1)^F \right) 
\left( 1+I_4 \right) + \frac{1}{4} \left( 1-(-1)^F \right) 
\left( 1-I_4 \right)
\eeqn
it appears, that the non BPS brane indeed only gives a 
``condensed'' description of the 
degrees of freedom of the brane anti-brane pair, all their Chan-Paton sectors being  
combined into a single one. But the traces, the projectors are inserted into, are in 
fact performed over different Hilbert spaces, differing in the momentum quantum 
numbers, not the oscillators.

\subsection{Zero mode spectra of T-duality invariant superpositions}

Later on we shall be particularly interested in  
D-branes of IIA invariant under the naive T-duality $(-1,1)^4$ 
of all four directions of the reflection $I_4$ and entirely wrapped on K3, 
and thus we look at either D04, D22$^\prime$ or D13 superpositions:
\beqn
\vert Dpp^\prime \rangle \equiv \frac{1}{\sqrt{2}} \left( 
\vert Dp \rangle + \vert Dp^\prime \rangle \right) = 
\frac{1}{\sqrt{2}} \left( 1 + (-1,1)^4 \right) \vert Dp \rangle .
\eeqn
Similar superpositions of D$p$- and 
D$(p+4)$-branes have already been encountered in asymmetric orientifolds, where in 
addition to an orbifold group quite similar to that used by KKS the world sheet 
parity was gauged \cite{hepth9812158,hepth9904092}. Fortunately, 
in the orientifold models there were no contributions to massless tadpoles in the 
sectors twisted by any asymmetric group element, and so 
there was no need of twisted branes. The question how 
twisted brane states could be defined, if they exist at all, and whether they 
allow any geometrical interpretation relating them to ordinary D-branes, maybe 
in the fashion of fractional branes, appears to be very 
challenging and involved. Indeed, a first step to construct such states has been taken 
in \cite{hepth9905024}. We shall eventually argue, that there again is no 
necessity to include them into our analysis of charges in the asymmetric orbifold. \\

The D$pp^\prime$ states above have four directions with 
mixed boundary conditions imposed on the open strings, therefore 
vanishing zero point energies in both sectors and a zero mode spectrum consisting of  
a single hypermultiplet of the effective theory in $d=6$, if no supersymmetry 
is broken by the background. The fermions in the R sector are invariant under 
the internal $SO(4)$ and those in the NS sector transform as ${\bf 2}_-$, due 
to the extra minus of $(-1)^F$ from the vacuum. 
Then the projection by $(1+{I}_4)$ keeps one half of the fermionic zero modes of 
the strings between the two branes, those of the R sector. 
The state leaves one quarter of the supersymmetry invariant and from the point of view 
of the effective theory in $d=6$ it breaks all supersymmetry in the orbifold background. 
This had to be suspected, as there are no quarter BPS multiplets in 
the perturbative spectrum of the dual heterotic model. 
The spectrum can also be derived directly from the closed 
string amplitude, the contribution of the entire T-duality invariant 
superposition of $\vert Dp\rangle \equiv \vert D(p,0,p) \rangle_{\mbox{\sz long}}$ 
states being:
\beqn  \label{above}
\int_0^\infty{dl\ \langle Dpp^\prime \vert e^{-lH_{\mbox{\sz cl}}} 
\vert Dpp^\prime \rangle} 
&=& \int_0^\infty{dl\ \left( \langle Dp \vert e^{-lH_{\mbox{\sz cl}}} 
\vert Dp \rangle + \langle Dp^\prime \vert 
e^{-lH_{\mbox{\sz cl}}} \vert Dp\rangle \right) } , \non
\int_0^\infty{dl\ \langle Dp^\prime \vert e^{-lH_{\mbox{\sz cl}}} 
\vert Dp\rangle} &=& 
\int_0^\infty{\frac{dt}{(2t)^{3/2}} 2^3 \left( \frac{f_3^4\left( e^{-\pi t} \right)
f_2^4\left( e^{-\pi t} \right) - f_2^4\left( e^{-\pi t} \right)f_3^4
\left( e^{-\pi t} \right) }{f_1^4\left( e^{-\pi t} \right)f_4^4
\left( e^{-\pi t} \right)} \right. } \non
& & \left. -
\frac{f_2^4\left( e^{-\pi t} \right)f_4^4\left( e^{-\pi t} \right)+
f_4^4\left( e^{-\pi t} \right)f_2^4\left( e^{-\pi t} \right)}{f_3^4
\left( e^{-\pi t} \right)f_1^4\left( e^{-\pi t} \right)} \right) \non
&=&  \int_0^\infty{\frac{dt}{2t} 
\Tr^{pp^\prime}_{\mbox{\sz NS-R}} \left( \frac{1}{4} \left( 1+(-1)^F \right) 
\left( 1+{I}_4 \right) e^{-2tH_{\mbox{\sz op}}} \right) } ,
\eeqn
the trace being performed only over the strings stretching between the two 
branes. The $f_i$ were 
defined in \cite{PC1987} originally, but we use again the convention of 
\cite{hepth9805019}, which differs by a single factor of $\sqrt{2}$ in $f_2$. 
The normalization now has to be chosen with prefactors 2 and $2i$ for the 
NSNS and RR components and equally for the untwisted and twisted components: 
$2^3 N_{\mbox{\sz U}}^2= 2^3 N_{\mbox{\sz T}}^2= {\mbox{Vol}}_{0+1}/(2\pi)$. 
The state $\vert Dpp^\prime \rangle$ may be $\vert D04 \rangle $ or 
$\vert D22^\prime \rangle$. The nonvanishing terms in the numerators originate from    
(NS-R), R$I_4$, NS$(-1)^FI_4$ in the open string trace, and obviously the NS 
zero modes cancel. The analogous amplitude for the 
non BPS states $\vert Dp\rangle \equiv \vert D(p,0,p) \rangle_{\mbox{\sz short}}$, 
e.g. $\vert D13 \rangle$ gives:
\beqn
\int_0^\infty{dl\ \langle Dp^\prime \vert e^{-lH_{\mbox{\sz cl}}} 
\vert Dp\rangle} &=& 
\int_0^\infty{\frac{dt}{(2t)^{3/2}} 2^3 \left( \frac{f_3^4\left( e^{-\pi t} \right)
f_2^4\left( e^{-\pi t} \right) - f_2^4\left( e^{-\pi t} \right)f_3^4
\left( e^{-\pi t} \right) }{f_1^4\left( e^{-\pi t} \right)f_4^4
\left( e^{-\pi t} \right)} \right. } \non
& & - \left. 
\frac{f_4^4\left( e^{-\pi t} \right)f_2^4\left( e^{-\pi t} \right)}
{f_1^4 \left( e^{-\pi t} \right)f_3^4\left( e^{-\pi t} \right)} \right)  \non
&=&  \int_0^\infty{\frac{dt}{2t} 
\Tr^{pp^\prime}_{\mbox{\sz NS-R}} \left( \frac{1}{2} \left( 1+(-1)^F I_4 \right) 
e^{-2tH_{\mbox{\sz op}}} \right) } ,
\eeqn
where only the R$I_4$ term is missing, as compared to (\ref{above}), and 
there are no tachyons, whatever relative signs or orientations are chosen.

\section{IIA on K3$\times S^1$ and the heterotic string on $T^5$}
\label{ch3}

In this section we like to review the spectra of states carrying particular charges on both 
sides of the duality IIA on K3$\times S^1$ to heterotic on $T^5$. 
The lightest among them are of special interest, as they are 
believed to be protected from quantum corrections in the fashion of BPS states 
in supersymmetric theories.

\subsection{Mapping of charges}

The way we like to figure the duality being considered 
first in \cite{hepth9410167,hepth9503124} relies very much on the results 
of \cite{hepth9504047}. We shall take the special point in the K3 
moduli space, where it can be written as an orbifold $T^4/I^4$ of the torus $T^4$ 
at self dual radii and also take the compactification torus of the 
heterotic side to be self dual, 
only then the asymmetric orbifold group is a symmetry of the theory.   
The duality map implies \cite{hepth9503124}
\beqn
g_{\mbox{\sz II}} = \frac{1}{g_{\mbox{\sz het}}},\quad 
r_{\mbox{\sz II}} = \frac{r_{\mbox{\sz het}}}{g_{\mbox{\sz het}}} .
\eeqn
In particular it involves the essential inversion of 
the coupling, which is not present in the fourdimensional version 
of the duality. The explicit matching of the degrees of freedom 
is achieved by a comparison of the zero modes of the 
IIA solitonic (``symmetric'') 5-brane \cite{CHS1991b}, compactified on the 
K3 down to $1+1$ dimension, and the world sheet 
fields of the weakly coupled heterotic string on the $T^4$. 
Both consist in eight chiral, say right moving, fermions, four bosonic 
fields for the translation zero modes in the noncompact directions and 
the internal bosons on a  
lattice $\Gamma_{4,20}$, stemming from the compactification torus 
respectively from the cohomology lattice. Their classical action is derived by dimensional 
reduction of the supergravity action in the soliton background and 
coincides with the classical world sheet action of the heterotic string 
\cite{hepth9504047}. 
Of course, apparently nothing is known about how to quantize the 5-brane. 
This twodimensional CFT may then be compactified further to $d=5$ 
by putting some of the fields on an additional circle.
In this spirit we shall later on apply the decomposition  
$({\bf 8}_\pm, \pm ) \rightarrow ({\bf 2}_\pm,{\bf 2}_\pm,\pm)$ 
according to $SO(9,1) \rightarrow SO(4) \times SO(4) \times SO(1,1)$ to the 
fermionic zero modes. The first (broken) $SO(4)$ acts on the K3, the second on 
$S^1 \times \mathbb{R}^3 \sim \mathbb{R}^4$ 
and the unbroken $SO(1,1)$ on the world sheet $\mathbb{R}^{1,1}$ 
of the dual string. The chirality invariant under the K3 projection is 
$({\bf 2}_+,...)$ and the chirality of the heterotic string world sheet 
fermions, respectively the Goldstone fields of the supercharges, whose symmetries are 
broken by the soliton, is $(...,+)$. \\

In general we take the entire lattice $\Gamma_{5,21}$ 
at some generic point in the moduli space, 
constrained as stated above. But to allow a separate identification 
of charges, we factorize it into
\beqn
\Gamma_{5,21} = \Gamma_{16} \oplus \Gamma_{4,4} \oplus \Gamma_{1,1}     
\eeqn
for the moment, which would correspond to an enhanced gauge group 
$SO(32) \times SO(8) \times SU(2)$. The duality can also be ``derived'' 
from a chain of S- and T-dualities, which at some point singles out 
one circle of the $T^4$, such that the winding and Kaluza-Klein (KK) 
momenta on it are related to the 0- and 4-forms, 
while those of the other directions of the $T^4$ are mapped to 
the 2-form charges. Recalling, that the duality mapping 
exchanges the field strength of the NSNS 
2-form with its dual and the vectors from the dimensionally reduced RR forms 
with the heterotic gauge fields form the internal lattice, 
we can easily write down the charges of the generic gauge group $SO(2)^{26}=U(1)^{26}$. 
Also we ignore, that the 3-form potential that enters the supegravity action 
is a combination of the RR forms and the NSNS 2-form. 
The heterotic momenta on the $\Gamma_{16}$ are then mapped to the sixteen 
D0-branes located at the fixed points, or in the blow up to the 
D2-branes wrapping the 2-cycles. At 
a generic point in moduli space these are not intersecting, and 
we shall have to turn on appropriate Wilson lines on the heterotic $T^4$ to break 
the gauge group accordingly. KK momenta and winding modes 
on the $T^4$ are related to D2-branes wrapping any of the six 2-cycles of 
the original torus of the K3, as well as to the D0- and D4-brane, the latter wrapping 
all of the K3, which together gives the $\Gamma_{4,4}$. 
Finally the winding and momenta of the heterotic string on the 
$S^1$ belong to the NS-5-brane of IIA wrapped entirely on K3$\times S^1$. 
In the following section we shall see, how this simple mapping gets refined, when 
deforming to a generic point in the moduli space. 
The magnetic charges on the heterotic side are all given as some excitation of 
the heterotic 5-brane, leaving just one extended spatial 
dimension after wrapping on 
four of the directions of the $T^5$. These are mapped to perturbative 
KK momenta and winding excitations of the elementary IIA string, 
when the fifth circle of the heterotic torus remained unwrapped, and to some 
D-branes, when any other circle is unwrapped. Thereby the magnetic 
dual of any D$p$-brane is a D$(p-6)$-brane, as usual, wrapped in order to leave 
one noncompact direction. While generically 
the radius $r_{\mbox{\sz II}}$ of the 
fifth circle on the type II side grows large in the limit of small $g_{\mbox{\sz het}}$, 
we could still consider any value by tuning the heterotic radius 
$r_{\mbox{\sz het}}$, and 
anyway the only BPS brane wrapping this circle is the NS-5-brane, whose mass by 
$r_{\mbox{\sz II}}/g^2_{\mbox{\sz II}} \rightarrow 
r_{\mbox{\sz het}}g_{\mbox{\sz het}}$ stays 
small, whereas any D-brane would acquire at least finite mass of the order of 
the compactification scale. 

\subsection{Spectra of BPS and non BPS charged states}

In the previous paragraph we have displayed all the charges in the two 
supersymmetric models, now we proceed by finding the multiplets of the 
lightest heterotic states that carry them at the generic point of the moduli space. 
It is pointed out, that a given D-brane of IIA corresponds to 
a combination of such charges on the heterotic side.   
We have seen in section \ref{ch2}, that there are no boundary states 
of the $I_4$ orbifold of IIA on flat space, that contained a twisted RR 
component and were extended in an odd 
number of dimensions inside the orbifold fixed plane. This continues to be 
valid when taking the compact version of this, IIA on a K3$\times S^1$, i.e. there are 
no RR charged BPS or non BPS boundary states for 
branes of finite volume, wrapping the $S^1$. At the 
orbifold limit of the K3 the boundary states defined in section \ref{ch2}  
are to be modified with the only respect that the momentum integration over the 
momenta on the $S^1$ direction has to be replaced by a sum over all KK
momenta, and that the strings may carry winding on those 
directions of the $T^4$ that are parallel to the brane, too. Such states have 
been discussed e.g. in \cite{hepth9805019,hepth9903123}, and the whole 
modification will be of no concern to us, 
as only the zero winding terms contribute to the massless spectrum, as long as we keep 
the $S^1$ at some generic radius at least. So we only need to regard 
D-branes on the K3, and these we have already examined in 
section \ref{ch2}. Their mapping to the heterotic states charged under 
the appropriate gauge fields has been explored in \cite{hepth9901014}, 
which we like to summarize briefly. For an original reference on toroidal 
compactifications of the heterotic string see \cite{G1987}. \\ 

In the perturbative heterotic spectrum precisely all states that have $N_R=c_R$, 
with $c_R$ denoting the zero point energies in the respective sectors, are 
BPS multiplets. On the right moving part they are vector supermultiplets
of $2^4=16$ states. The lightest states carrying any momentum 
$V^2=2$ on the internal $\Gamma_{16}$ transform in this multiplet, 
while a BPS multiplet carrying only KK momentum on the $\Gamma_{4,4}$ 
necessarily starts with a copy of the entire massles multiplet. The 
Wilson lines that break the gauge group introduce a mixing of charges, as the winding 
states automatically have $V^2 > 0$. The charges of IIA D-branes are translated as follows: 
The bulk BPS D0-brane refers to 
heterotic states carrying KK momentum only on the special direction of the $T^4$, 
that was related to the D0- and D4-brane part of the cohomology lattice. 
The fractional BPS D0-brane is also charged under one of the twisted sector 
gauge fields in addition to the KK 
momentum of the bulk D0-brane. The BPS D4-branes are mapped to the winding modes 
in the special direction and additionally have got 
a nonvanishing momentum on the $\Gamma_{16}$, too. 
Finally the BPS D2-branes belong to the winding and momenta on the 
remaining three directions of the torus, momentum and winding being distinguished 
by the criterion, if the brane wraps the special circle or not, and again they 
carry twisted sector charge. To summarize: The only BPS charge, 
whose lightest states make up a gravity multiplet plus vectors, is that of the bulk 
D0-brane, while all other charges appear in vector multiplets on the 
heterotic side. \\

As well an example of a non BPS multiplet has been discussed. 
In the right moving heterotic spectrum the lightest non BPS $(N_R=c_R+1)$ multiplet is 
\beqn \label{longmult}
\psi_{-3/2}^\mu \vert 0 \rangle_{\mbox{\sz NS}},\ 
\psi_{-1/2}^{\left[ \mu \right.} \psi_{-1/2}^\nu 
\psi_{-1/2}^{\left. \rho \right]} \vert 0 \rangle_{\mbox{\sz NS}},\ 
\psi_{-1/2}^\mu \tilde{\alpha}_{-1}^\nu \vert 0 \rangle_{\mbox{\sz NS}}\ 
\oplus\ \psi_{-1}^\mu \vert 0 \rangle_{\mbox{\sz R}},\ 
\tilde{\alpha}_{-1}^\mu \vert 0 \rangle_{\mbox{\sz R}} ,
\eeqn
and it may have $V^2=4$ but no winding or KK momenta on the $T^4$. This can be realized 
on the IIA side by a superposition of a D0-brane and an 
anti-D0-brane at different or at the same fixed points. Taken separately, they are 
mapped to states with $V^2=2$ and momenta on 
the $T^4$, that cancel each other. It leads to the projection 
\beqn
P= \frac{1}{8} \left( 1-(-1)^F \right) \left( 1-I_4 \right) \left(1+s \right)
\eeqn
in the open string spectrum of the strings stretching between the different 
branes, where $s$ acts as a shift of the position by one half 
of the radius of the torus. As the projection without the shift already 
removes the tachyon from a 
D0-brane (no extension transverse to the fixed plane), the state will be stable. 
An alternative description of this state as a non BPS 
D1-brane was given \cite{hepth9812031,hepth9901014}. It stretches between the 
two fixed points and carries a NSNS component in the untwisted sector and 
two twisted sector RR components. The open string spectrum has the projection   
\beqn
P= \frac{1}{4} \left( 1+(-1)^F I_4\right) \left(1+s \right) .
\eeqn
As the D1-brane 
has its spatial dimension transverse to the orbifold fixed plane, it will not have 
the open string NS groundstate projected out in general, the lightest potentially 
tachyonic state being the one with minimal allowed momentum in this direction.

\subsection{The dual pair of asymmetric orbifolds} 

While the models that were considered by KKS and others were 
defined in $d=4$, it was necessary to decompactify at least one dimension, 
to be able to apply a duality from IIA to the heterotic string, which 
inverts the coupling. On the other hand it is crucial to have still 
one extra circle in addition to the K3, as only by the shifts in 
this coordinate the action of the orbifold becomes free of fixed points and 
the adiabatic argument applicable \cite{hepth9507050}. We shall focus on models 
of the kind of that 
constructed by Harvey \cite{hepth9807213}, where the orbifold group is generated 
by two elements $f$ and $g$ with $fg=I_4$, 
acting on the $T^5$, which we only need schematically:
\beqn
f \sim (-1,1)^4(-1)^{F_R}\ {\mbox{and shifts}}, \quad 
g \sim (1,-1)^4(-1)^{F_L}\ {\mbox{and shifts}}.   
 \eeqn
It is found, that one has some freedom in arranging the shifts, which are 
constrained in particular by the requirement of level matching, and by 
$fg=I_4$. The orbifold 
projection completely eliminates supersymmetry and leaves Bose Fermi matching 
perturbatively, as for the KKS models, at least for the one-loop level. 
The group can be split off, such one can 
divide by $I_4$ first and only then by an asymmetric element, say $f$. 
After using the duality of IIA on $T^4/I_4 \times S^1$ to heterotic on $T^5$ in the 
first step, one deduces the action $f_{\mbox{\sz het}}$ of $f$ on the 
heterotic side, and as it is free, 
hopes for the duality to be inherited by the pair of nonsupersymmetric orbifolds:
\beqn
f_{\mbox{\sz het}} \sim \left( (-1)^8, 1^8 \right)_I (-1,1)_i^4 (-1)^{F_R} 
\quad \mbox{and shifts.}
\eeqn
There may appear to be some ambiguity in how to generalize the transformation 
properties under $f_{\mbox{\sz het}}$ from the massless fields to the massive. 
On the IIA side the zero modes of the 5-brane give the world sheet field theory 
of the heterotic string after compactification on K3. Their transformation 
under $f$ can directly be deduced from the tendimensional massless fields, and 
it is just the action we have given above: One half of the RR forms get 
reflected and only the NSNS fields with no component in the $I_4$ directions 
are invariant.  
Then by duality this has to be the action on the heterotic world sheet 
fields without anything left to be guessed, except for the shifts. \\

In the twisted sector of the heterotic orbifold 
there are no massless fermions anyway, due to the positive zero point energy, and the 
only boson of potentially zero mass is the NS ground state, whose 
mass depends on $r_{\mbox{\sz het}}$, the tachyonic region, depending on the 
nature of the shift on the $\Gamma_{1,1}$, at large or small radius. In particular 
the twisted sector does not display any new massless vector fields, such that 
there are no new gauge degrees of freedom and no new charges present. Correspondingly 
there should be no massless RR forms in the twisted sector of the dual IIA 
orbifold showing up and therefore no twisted sector D-branes or boundary 
states to be regarded. 
This argument is surely not entirely satisfactory and should be accompanied by an explicit 
construction of the twisted sector boundary states or a proof of their absence. 
Unfortunately we do not know a way to do this presently. \\ 

Let us also note the action of $f_{\mbox{\sz het}}$ on the multiplets with $(N_L=1,N_R=c_R)$ 
and $(N_L=0,N_R=c_R+1)$ explicitly. Any 
of the lightest state carrying a given RR charge transforms in one of these (or 
a BPS vector multiplet $(N_L=0,N_R=c_R)$). We assume to 
have chosen a combination of charges that 
is invariant under the action of $f_{\mbox{\sz het}}$ on the lattice, so there 
will be no further signs to be considered. In the first case  
one half of the gauge degrees of freedom are invariant 
and of the graviton multiplet all the gravitinos and one 
half of the vectors are odd. This leaves Bose Fermi matching in the massless spectrum 
and even a supermultiplet for the diagonal of the gauge group.  
The second is the lightest non BPS multiplet. The projection onto 
invariant states simply removes all fermions and leaves all bosons. 
We immediately recognize the violation of Bose Fermi matching at the massive 
level of the untwisted sector. In particular it is manifest in non BPS multiplets 
that will be found to be 
the lightest to carry some charge under the gauge group in the orbifold.

\section{Charged states on the asymmetric orbifolds}
\label{ch4}

In the preceding sections we have collected a couple of facts about the 
spectra of states carrying some particular charges under the gauge group on 
both sides of the (conjectured) 
duality. 
On the heterotic side these are perturbative states of the untwisted sector, 
invariant under the orbifold projection. On the IIA side the charges map to  
superpositions of D-branes. Their zero modes should, after quantization on the 
worldsheet of the dual string, reproduce at least the lightest heterotic states of 
a given charge. While we do not have a precise understanding, how the 
supergravity solution should be modified on the asymmetric orbifold, we have been able to 
deduce the zero mode spectra of D-branes from the according boundary states. 
We shall now figure out, which space-time fields they would generate as their lightest 
excitation, if 
taken to be world sheet fields of the dual string, subject to the asymmetric orbifold 
projection, and compare to the spectra of non BPS states on the heterotic side.

\subsection{Fractional BPS branes}

We first look at the boundary states 
of the K3$\times S^1$ orbifold, as derived in section \ref{ch2}, and 
afterwards we come to the invariant superpositions of branes in the untwisted sector 
of the asymmetric orbifold. Branes of the twisted sector do not need to be inspected, as 
there are no additional charges arising, due to the absence of further gauge fields.   
Recall first, that the BPS soliton of IIA breaks the supersymmetries 
\beqn
({\bf 2}_+,{\bf 2}_\pm,+),\ ({\bf 2}_-,{\bf 2}_\pm,-)
\eeqn
of even chirality on its worldvolume \cite{CHS1991b}, 
referring to the zero modes kept 
by the usual GSO projection of the open strings ending on a BPS bulk brane. 
The K3 leaves all spinors $({\bf 2}_+,...)$ invariant, i.e. those 
that are even under $I_4$. The zero modes on the fractional 
branes that are invariant under the orbifold projection  
in the open string sector via (\ref{projfrac}) are then given by   
those which are unbroken by the K3 and even under $(-1)^F$:
\beqn
({\bf 2}_+,{\bf 2}_\pm,+) .
\eeqn 
They are both of positive chirality on the world sheet, identified with the right moving 
world sheet spinors of the heterotic theory. After the quantization 
of the world sheet theory of the dual string they generate a ground state degeneracy 
of $2^{8/2}=16$ and reproduce as their lowest excitation the 
tendimensional BPS vector multiplet, which the lightest states carrying the charge 
of the D-brane in the perturbative heterotic spectrum transform in \cite{hepth9504047}. 
Also we notice, that the heavier states of the heterotic string, carrying the 
same charge, would involve excitations of all left moving bosonic oscillators 
on the world sheet, 
whereas the zero mode spectrum of the D-brane involves only the 
bosonic Goldstone fields for the translations transverse to the K3, not enough 
to reproduce the heterotic multiplets.

\subsection{Non BPS branes}

The more interesting case is that of the non BPS brane with nonsupersymmetric 
projection (\ref{projnbps}) in the open string spectrum. In addition to the zero 
modes of the fractional brane from open strings with GSO projection 
there are those, which are odd under 
$(-1)^F$ and $I_4$, therefore belong to supersymmetries that 
remained unbroken by the BPS soliton and transform 
as $({\bf 2}_-,...)$, together:
\beqn  
({\bf 2}_+,{\bf 2}_\pm,+),\ ({\bf 2}_-,{\bf 2}_\pm,+) .
\eeqn
These sixteen zero modes are still chiral on the world sheet, 
as they better had to, and upon quantization they generate a 
degeneracy of $2^{16/2}=256$ on the right moving side, which, subject to the GSO 
projection in $d=10$, gives rise to 16 fermions, half of them even, the other half odd 
under $I_4$. This looks exactly like the fermionic states  
as given in (\ref{longmult}) for the lightest non BPS multiplets of the 
heterotic string with $V^2=4$. 
Again the more massive states are not reproduced correctly by the single D-brane, 
as not enough bosons are available. Up to now we have omitted the  
tachyon that may be present in the open string spectrum on the IIA side. 
In fact it will be there, as in general the branes 
are extended on the K3. But the open string sector with momenta 
odd under $I_4$ does not have any additonal zero modes, 
except at the special radius, when the tachyon gets massless. 
The deformation that is associated with this additional bosonic field was shown, 
even if only in a quite different situation, to allow to pass to an equivalent 
theory in a lower dimension but with a tachyon condensate 
\cite{hepth9808141,hepth9805170}. If this mechanism works 
in our case, it will reduce or enlarge the number of dimensions transverse to the 
orbifold fixed plane, acquiring some condensate, which only neutralizes 
the energy density of the brane in these directions, and finally the above argument 
goes through without any additional zero modes to be regarded. In the 
present case for instance a D1-brane stretching between two 
fixed points gets deformed to a pair of D0-branes, brane and anti-brane, 
sitting at the two fixed points. The oscillator parts of the 
open string Hilbert spaces only differ by states, which had 
combinations of momenta on the D1-brane, that were odd under $I_4$. Therefore the 
zero modes of the stable brane anti-brane configuration with tachyon condensate 
are just those of the unstable non BPS brane and the spectra of charged states identical. 
In this sense (\ref{gleich}) really gets to work after tachyon condensation.  
In the supersymmetric models the bosonic states should then follow automatically, indeed.

\subsection{Branes of the asymmetric orbifold}

Only little remains to do the same analysis for the states of the untwisted 
sector of the asymmetric orbifold. Following our discussion of section \ref{ch3}, 
it acts on the entire $\Gamma_{4,20}$ by a 
reflection in the sense, that the momenta on the $\Gamma_{16}$ are exchanged 
in some fashion and the winding and KK momenta on the $\Gamma_{4,4}$ also. 
On the type II side this amounts to mapping fractional D0-branes to D4-branes 
with Wilson lines given by the image of the charge of the D0-brane on the $\Gamma_{16}$. 
D2-branes wrapping a 2-cycle of the $\Gamma_{4,4}$ are mapped to the D2-branes 
wrapping the naively dual 2-cycle in the other two directions, and their 
twisted charges on the $\Gamma_{16}$ also get exchanged. 
But the obvious invariant combinations, a brane and its image, 
are not the lightest states to carry any given charge. Such an 
invariant D22$^\prime$-brane superposition would carry winding and momentum 
on the $T^4$ as well as 
internal momentum with respect to two of the $SO(2)$ from the $\Gamma_{16}$ in 
the heterotic picture. But, of course, there are also heterotic states 
which carry invariant combinations $V^2=(v_1+v_2)^2=4$ 
of internal momenta, but their winding and momenta on the $T^4$  
vanish. These are mapped to brane anti-brane superpositions, the lightest 
states to carry any orbifold invariant charge on the $\Gamma_{16}$. 
The D0-brane anti-brane pair, 
respectively the presumably equivalent non BPS D1-brane, has 
already been analyzed above as a non BPS state, which carried the fermionic 
zero modes to generate the space-time fermions of (\ref{longmult}), the 
heterotic non BPS multiplet, and was stable for some complementary 
regions of the radii. The orbifold projection of IIA applied on the world sheet 
of the dual string now gives a minus to all fermions, that 
are left moving on the world sheet and odd under $I_4$ by $(-1,1)^4$, i.e. to 
none, and a minus to all, that are right moving by $(-1)^{F_R}$, 
i.e. to all. All the fermions are removed and only bosons left, the same action as  
on the heterotic non BPS multiplet, which was a source of the 
violation of the matching of bosons and fermions. Note, that we used the 
orbifold projection on the IIA side only to deduce this and did not 
require any information on the heterotic translation $f_{\mbox{\sz het}}$. 
While this approach gives an interpretation of how the mismatch emerges 
in the nonperturbative regime of the IIA orbifold, it surely 
does not allow to do any quantitative computation of the contributions 
to the partition function. \\

There are several other cases left, one has to consider, for instance the 
superposition of a D4- or D2-brane with an anti-brane wrapping 
the same directions of the torus. The lightest heterotic oscillator 
excitations are again given by (\ref{longmult}) and the 
derivation of the zero modes proceeds as above. \\

One might think, that there also were states carrying equal winding 
and KK momentum quantum number on the $T^4$ but no twisted sector 
charges. In fact no such superpositions without momentum on the $\Gamma_{16}$ 
can occur, as the presence of Wilson lines enforces $V^2 >0$ for winding states. 
The application of $f_{\mbox{\sz het}}$ does not map the BPS states with KK 
momentum and $V^2=v_1^2=2$ to the BPS winding states with $V^2=(v_2+A)^2=2$,  
$A$ indicating the Wilson line in the direction of the winding, for 
instance as defined in \cite{hepth9901014}. Also one needs to notice, that the left handed 
reflection enforces winding equal to minus the canonical momentum, as opposed 
to a right handed reflection. There are in fact invariant states in the heterotic 
spectrum with opposite momentum and winding as well as $V^2=(v_1+v_2+A)^2=4$. The 
level matching demands for the lightest such states $N_L=0$ and $N_R=c_R$, which before 
orbifolding is the BPS vector multiplet. The obvious candidate for an invariant  
boundary state to describe the dual state 
with the required RR charges is a superposition of a D2-brane 
and an anti-D2-brane wrapping disjoint circles of the K3, according to opposite 
momentum and winding. It is one of the stable superpositions with mixed 
boundary conditions in four directions, which  
we have discussed in section \ref{ch2}, with the exception, that  
in the Chan-Paton sector of open strings between the two branes the projector is 
\beqn
P= \frac{1}{4} \left( 1- (-1)^F \right) \left( 1- I_4 \right),  
\eeqn 
but again no tachyons can appear. In the other Chan-Paton 
sectors it has the fermionic zero modes of a fractional brane, whereas   
in the mixed sector all zero modes from the R and NS sectors now are projected out.
The resulting zero mode spectrum is that of a fractional D0-brane, which again 
generates a chiral world sheet theory on the dual string with appropriate degeneracy 
of the vacuum to reproduce the BPS vector multiplet as its lowest excitation. 
The orbifold projection then removes the fermions and leaves the 
bosons, as before. \\

The D22$^\prime$ 
superposition is presumably equivalent via tachyon condensation 
to an instable superposition of a D3- and a D1-brane both stretching 
between the two fixed points and with opposite bulk charges. 
In this spirit D0- anti-D4-brane superpositions and their tachyonic equivalent, 
a D1- anti-D3-brane superposition wrapping the special circle, can also be treated.

\section{Summary}

We have analyzed the spectra of boundary states that arise in the untwisted 
sector of a particular class of asymmetric orbifolds. They were found to be 
given by superpositions of branes of the original IIA theory on K3, that are 
invariant under a version of T-duality. The lightest states carrying a given 
conserved charge were found to consist of brane anti-brane pairs. 
For any such pair there was an alternative description in terms of non BPS 
branes, the equivalence defined via the condensation of the 
tachyon arising in the open string spectrum. Assuming 
the validity of the adiabatic argument and the continuity of some charged non BPS 
states, the zero modes of the open strings on these branes were shown to be chiral 
on the world sheet of the dual string and to 
be candidates to generate a spectrum of space-time fermions that have the correct 
multiplicity and the same transformation properties under the orbifold reflection $I_4$, 
as the lightest heterotic states that carry the same charge. This may be figured to be a 
consistency check of the applicability of the adiabatic argument and the underlying 
duality and as a suggestion, how the mismatch of bosons and fermions in the 
nonperturbative regime of the IIA orbifold appears. \\

\noindent {\bf Acknowledgements} \\

I like to thank A. Krause, A. Miemiec and in particular R. Blumenhagen 
for valuable discussion, as well as M. Gaberdiel for answering a question of mine.

\bibliography{books,reviews,articles}
\bibliographystyle{unsrt}

\end{document}